\documentclass{jetpl}
\twocolumn
\usepackage{cite}
\usepackage{graphicx}
\usepackage{epstopdf}
\lat

\title{Oscillations of the Critical Temperature in a (Fe/Cr/Fe)/V/Fe Heterostructure}

\rtitle{Oscillations of $T_c$ in a (Fe/Cr/Fe)/V/Fe Heterostructure}

\sodtitle{Oscillations of the critical temperature in a (Fe/Cr/Fe)/V/Fe heterostructure}

\author{V.\,A.\,Tumanov$^{a}$ \/\thanks{e-mail: tumanvadim@yandex.ru},
Yu.\,V.\,Goryunov$^{b}$, Yu.\,N.\,Proshin$^{a}$}

\rauthor{V.\,A.\,Tumanov, Yu.\,V.\,Goryunov, Yu.\,N.\,Proshin}

\sodauthor{Tumanov, Goryunov, Proshin}

\address{$^a$Kazan Federal University, Kazan, 420008 Russia\\~\\
$^b$Zavoisky Physical-Technical Institute, FRC Kazan Scientific Center,
Russian Academy of Sciences, Kazan, 420029 Russia}

\abstract{The superconducting and magnetic properties of the (Fe/Cr/Fe)/V/Fe layered system with variable thickness of the chromium layer have been experimentally and theoretically studied. The magnetic properties of the system have been studied by the ferromagnetic resonance method, and the superconducting transition temperature has been measured from the jump in the magnetic susceptibility. A wide variety of magnetic states are observed in the system; in particular, the structure of small domains can arise in the iron layer placed between	vanadium and chromium. It has been shown experimentally that the critical temperature $T_c$ of the superconducting	transition undergoes nonmonotonic oscillations with a noticeable amplitude in the given system with the change in the thickness of the Cr layer. The proposed model based on the proximity effect theory makes it possible to relate these $T_c$ oscillations to the features of the magnetic structure of the samples.}

\begin{document}

\maketitle

The coexistence of superconductivity and ferromagnetism within a uniform sample requires specific conditions, which are difficult to fulfill. It can be achieved in superconductor–ferromagnet heterostructures either by spatial separation of ferromagnetic and superconducting materials (the proximity effect) or by the suppression of the effective exchange field. The spatial separation makes it possible to combine magnetic and superconducting properties within a single sample because of the large delocalization of Cooper pairs, which transfer superconducting correlations in the near-boundary layer of a ferromagnet. The emerging competition of superconductivity and magnetism leads to the appearance of a number of interesting effects, in particular, a nonmonotonic dependence of the critical temperature and Josephson current on the thicknesses of the ferromagnetic layers $d_f$ (see reviews \cite{bib:Buzdin05,bib:Efetov08,bib:Izyumov02,bib:Soloviev17} and references therein). Magnetic inhomogeneities of various nature (domain walls, helicoidal magnetic structures, artificially created multilayer structures with different directions of magnetization) in the ferromagnetic layer significantly complicate the structure of superconducting correlations
in the superconductor–ferromagnet system. The directionally inhomogeneous magnetization within
the ferromagnetic layer leads to the appearance of triplet superconducting correlations with a nonzero spin projection \cite{bib:Fominov07,bib:Khaire10,bib:Buzdin11} and affects the critical temperature of the superconducting transition. Theoretical estimates of this effect were carried out, e.g., in \cite{bib:Houzet06,bib:Linder14}. 

In this work, we study the magnetic and superconducting properties of a system where it is possible to form domains whose dimensions are about the superconducting coherence length. The experimental implementation of such a system is the contact of a superconductor with a magnetic system (Fe/Cr/Fe), which in itself is of great interest \cite{bib:Unguris_PRL,bib:Wang90}. In the (Fe/Cr/Fe) system, the phenomenon of giant magnetoresistance is observed \cite{bib:Zahan95}, while the mutual orientation of the magnetizations of the iron layers as a function of the thickness of the chromium layer is very complex and depends on the conditions for the deposition of the layers \cite{bib:Unguris_PRL}.

\begin{figure}
	\includegraphics[width=1 \linewidth]{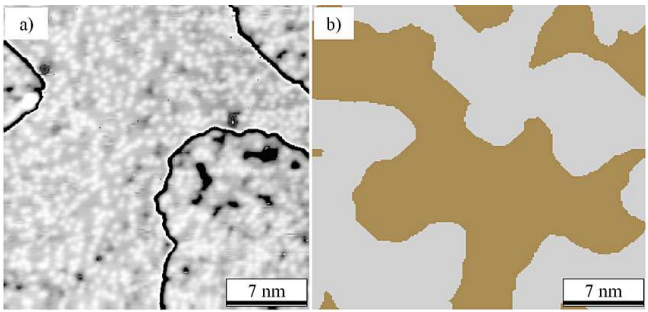}
	\caption{\textbf{Fig.\,1.} (Color online) (a) Scanning electron micrograph
		demonstrating chromium islands on the iron surface. The
		effective thickness of chromium is 0.4 monolayers. The
		figure is taken from \cite{bib:Davies_PRL}. (b) Simulation of chromium
		deposition on the surface of iron with a half-filled top layer
		within our phenomenological model. The effective thickness
		of chromium is 0.43 monolayers. }
	\label{GR}    
\end{figure}

For the studies, two series of samples were prepared on a single-crystal MgO (001) substrate. The first series of Fe/V(335\,\AA)/Fe samples include symmetric wedge-shaped iron layers, where the top layer is protected by Pd (20\,\AA). The second series of Fe\thinspace(8\,\AA)/Cr/Fe\thinspace(8\,\AA)/V\thinspace(340\,\AA)/Fe\thinspace(20\,\AA) samples have a wedge-shaped chromium layer, where the upper layer is protected by vanadium (60\,\AA). Here and below, the layers are listed from left to right in the order of deposition. During the deposition, the temperature of the substrates was  $300^{\circ}$C, which is optimal for obtaining the smoothest layers and, correspondingly, sharp interfaces \cite{bib:Unguris_PRL,bib:Romanovskiy16,bib:Garifullin_PRB}. The process of obtaining samples is described in detail in \cite{bib:Garifullin_PRB,bib:Goryunov07}. We note that an iron layer 8\,\AA\ thick in the Fe/Cr/Fe magnetic system corresponds to 5.5 monolayers. At this thickness, the upper iron monolayer in the first layer is half completed even in the case of a substrate with zero roughness. The upper monolayer in the first iron layer is not continuous and is a set of islands. The islands of chromium growing on iron were experimentally studied in \cite{bib:Davies_PRL} by the scanning tunneling microscopy method. The characteristic size of the observed islands was 100--250\,\AA\  (Fig\,\ref{GR}a). The growth of such islands can be reproduced within a qualitative model of the epitaxial deposition in which the stability of the position of the atom in the deposited layer is determined by the number of nearest neighbors. Figure\thinspace\ref{GR}b shows the results of our calculations within this model.

The magnetic properties of the (Fe/Cr/Fe)/V/Fe samples with varying thickness of the chromium layer were studied by the ferromagnetic resonance (FMR) method. The signal from the Fe/Cr/Fe fragment can be separated from the signal of the top iron layer with a thickness of 20\,\AA. In the absence of a Cr layer, these properties are very close to those of a signal of the sample with a Fe layer thickness of 16\,\AA. When this layer is divided into two Fe layers each 8\,\AA\ thick by the introduction of the first chromium monolayer, the FMR signal properties change significantly. On one hand, these changes are caused by a decrease in the thickness of the Fe layer and, correspondingly, by averaging the Neel contribution to half the thickness. On the other hand, these properties are affected by the magnetization of the chromium layer. A detailed analysis of the FMR spectrum of a sample of the (Fe/Cr/Fe)/V/Fe series with the effective thickness of the chromium layer less than one monolayer was performed in \cite{bib:Goryunov07}. The form of the angular dependence for the samples with a thicker chromium layer is insignificantly different, but the relative intensity of the FMR signal of the Fe/Cr/Fe fragment is much lower. This indicates that a large part of the sample with the chromium layer about 2--3\,\AA\ in thickness ceases to be ferromagnetic.

\begin{figure}
	\centering
	\includegraphics[width=0.95 \linewidth]{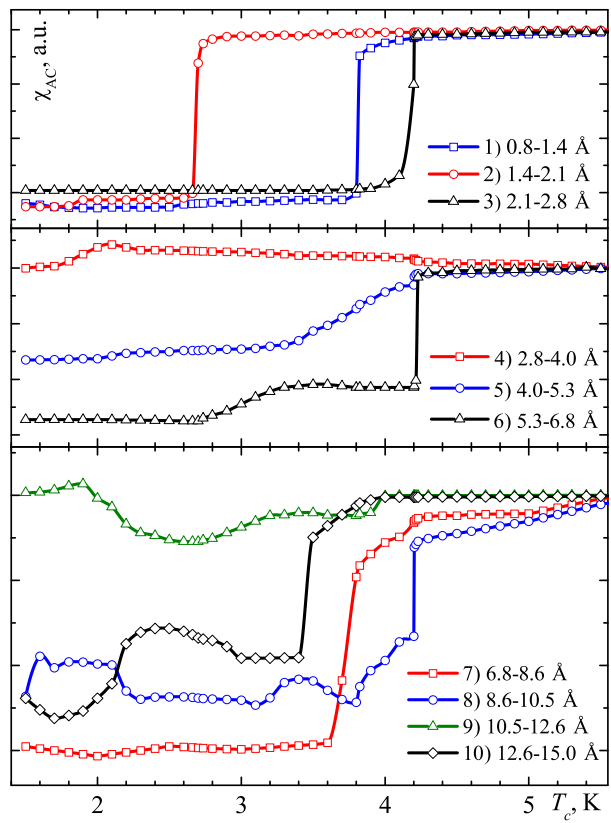}
	\caption{\textbf{Fig.\,2.} (Color online) Superconducting transitions in the
		(Fe/Cr/Fe)/V/Fe system measured from the jump in the
		magnetic susceptibility.}
	\label{AC}   
\end{figure}

\begin{figure}
	\includegraphics[width=1 \linewidth]{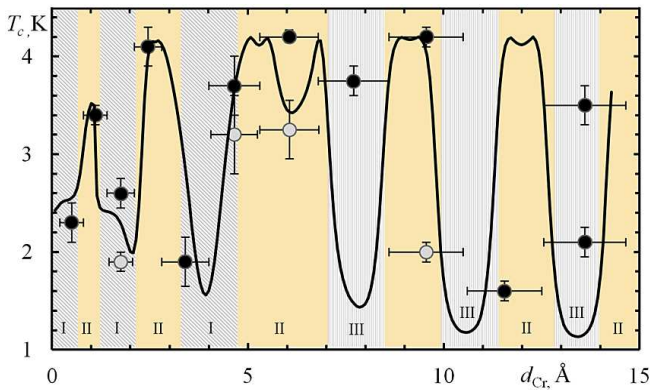}
	\caption{\textbf{Fig.\,3.} (Color online) Critical temperature of the superconducting
		transition of the (Fe/Cr/Fe)/V/Fe system obtained from the experimental magnetic susceptibility
		data: the solid circles correspond to the transition temperature of the majority of the sample, the empty circles correspond to the weaker transitions in the same samples,	and the solid line is the theoretically estimated dependence	of the critical temperature on the thickness of the chromium layer. The thickness ranges of region I correspond to	parallel magnetizations of iron layers separated by a chromium	layer; region III, to antiparallel magnetization; and
		region II, to a shallow domain structure in the iron layer adjacent to the superconductor.}
	\label{Tc}     
\end{figure}

The temperature of the transition to the superconducting state was determined by the jump in the magnetic
susceptibility at the alternating current. This method determines the relative magnitude of the superconducting phase in the sample. Figure\thinspace\ref{AC} shows the results of measurements of the magnetic susceptibility for a series of (Fe/Cr/Fe)/V/Fe samples and the thickness ranges of the chromium layer corresponding to all samples. In view of the wedge shape of the Cr layer, its thickness within one sample varies by about one monolayer, and we can observe that different sections of the sample become superconducting at different temperatures. Particularly, we note sharp transitions at high temperatures of 4.1--4.25\,K observed in samples 3, 6, and 8. These transitions are supplemented by weak wide transitions at a lower temperature. Samples 2, 4, 8, 9, and 10 demonstrate weak transitions near 2\,K. The dependence of the critical temperature of the superconducting transition on the thickness of the chromium layer is shown in \,\ref{Tc} along with our theoretical estimate.

A chromium monolayer grown on a thin iron layer is ferromagnetically ordered within the layer. In this
case, the magnetization of the chromium layer is antiparallel to the magnetization of the iron layer [17]. Thus, in the case of the chromium growth on a filled iron layer, when the iron layer is divided by an odd (even) number of chromium monolayers, a parallel (antiparallel) orientation of the magnetizations in adjacent iron layers can be expected. However, for a sufficiently small thickness of the chromium layer between the ferromagnetic layers, an RKKY-type interaction arises \cite{bib:Wang90}, which has the character of
damped oscillations with a much longer period. The roughness of the iron layer also introduces significant corrections, which should be taken into account when analyzing the experimental results. The sharpest boundaries are reached at a deposition temperature of  $300^{\circ}$C. In the experiment, with a change in the thickness of chromium \cite{bib:Unguris_PRL}, a complex dependence of the mutual orientation of the magnetization in the iron layers is observed with combination of long and short oscillation periods. Long-period oscillations appear up to thicknesses of 10--15\,\AA, after which they pass into oscillations with a period approximately equal to the lattice constant of chromium (see Fig. 4 in \cite{bib:Unguris_PRL}).

 To analyze systems in which the magnetization in the plane of the boundary changes slowly on the scales of the coherence length, it suffices to analyze the behavior of the pair amplitude perpendicular to the plane of the layers. Taking into account the estimates in \cite{bib:Garifullin_PRB} for the mean free path and the spin stiffness length in such systems, we perform calculations within the dirty limit and use the Usadel equations. To obtain
semiquantitative estimates, we will use the simplest single-mode approximation, which proved to be effective for systems with a large number of layers \cite{bib:Radovic88}. In this approximation, the pair amplitude in the superconducting layer is sought in the form of a harmonic function, the wave vector in which can be found from the generalized Kupriyanov–Lukichev boundary conditions \cite{bib:Kupriyanov82}. The wave vector in ferromagnetic layers is related to the effective exchange field by the relation $k_{f}=\sqrt{-2iI/D_f}$. We neglect changes in the pair amplitude within a thin layer of the antiferromagnetic metal. In this case, the antiferromagnetic layer is replaced by an interface between the ferromagnetic layers, the matching conditions on which have the form 
\begin{equation}
\dfrac{d}{dx} F_{f1} =\dfrac{d}{dx} F_{f2} =\frac{(F_{f2} - F_{f1})}{2\sigma_\mathrm{Fe/Cr}R_\mathrm{Fe/Cr}},\\
\label{eq:boundary_cond}
\end{equation}
where $F_{f1}$ and $F_{f2}$ are the singlet parts of the pair amplitude in ferromagnetic layers, $\sigma_\mathrm{Fe/Cr}$ is the parameter characterizing the transparency of the Fe/Cr interface \cite{bib:Buzdin05}, $R_\mathrm{Fe/Cr}$ and is the resistance per unit area of the Fe/Cr interface. In the Fe/V/Fe system, pronounced oscillations of the critical temperature as a function of the thickness of the ferromagnetic layers are observed \cite{bib:Garifullin_PRB}. The dependence of the critical temperature on the thickness of the ferromagnetic layer has a deep minimum and a plateau of 0.4$T_{cs}$, where $T_{cs}$ is the critical temperature of the superconducting transition in the bulk superconductor. This behavior of the critical temperature was studied in numerous theoretical works \cite{bib:Buzdin05,bib:Efetov08,bib:Izyumov02,bib:Avdeev13,bib:Koshina17} and makes it possible to determine, by fitting, certain parameters of the theory of the proximity effect. We used the experimental values of the superconducting coherence length $\xi_s$\,=\,125\,\AA, BCS coherence length $\xi_{BCS}$\,=\,440\,\AA\, and the ratio of the mean free path to the spin stiffness length $l_f/\xi_f$\,=\,1.3 from \cite{bib:Garifullin_PRB}. The spin stiffness length $\xi_f$\,=\,11.5\,\AA,   the transparency parameter of the V/Fe interface on the vanadium side $\sigma_{V/Fe}$\,=\,7.5, and the parameter $n_{sf}=(N_fv_f)/(N_sv_s)$\,=\,0.23 ($N_{s(f)}$ and $v_s(f)$ are the density of states and the velocity at the Fermi level, respectively) were determined from fitting the theory to experimental data \cite{bib:Garifullin_PRB}, where the same materials and methods of sputtering were used. The parameter $p=(\sigma_\mathrm{Fe/Cr}R_\mathrm{Fe/Cr})/(\sigma_\mathrm{Fe/V}R_\mathrm{Fe/V})$ characterizing the transparency of the Fe/Cr interface was used as a fitting parameter when comparing our theory with experiment. We took into account that the transparency parameter depends on the mutual orientation of the magnetization of the iron layers in the Fe/Cr/Fe system because of the giant magnetoresistance (GMR) phenomenon and we used the results from \cite{bib:Zahan95,bib:Romanovskiy16}. 

The analysis of the system containing the domain structure in the layer closest to the superconductor requires a theory including the spatial inhomogeneity of the Usadel function in the plane of the interface. For simplicity, we consider only an effective two-layer system. As the upper layer, we use the system V\,(340\,\AA)/Fe\thinspace(20\,\AA), which in this case serves as a superconducting layer with  $T_{cs} \simeq 4.3 $\,K. This is possible, because the 20\,\AA\ iron layer corresponds to the dependence $T_c(d_{Fe})$ reaching the plateau. As the lower layer, we consider the iron layer between vanadium and chromium, with the interface with chromium replaced by a free boundary, because according to our estimates, its transparency is much lower than the transparency of the Fe/V interface. In order to obtain an analytical solution of the boundary value problem, we consider the case of thin ferromagnetic and superconducting layers. In this approximation, the real parameters of the system (diffusion coefficient $D_{s(f)}$, exchange field $I$, and superconducting order parameter $\Delta$) are replaced by effective parameters averaged
with allowance for the boundary conditions along the axis perpendicular to the plane of the boundary \cite{bib:Kulic01}:
\begin{equation}
\label{effective}
\begin{array}{l}
{\eta _{s(f)}} = \frac{{{\sigma _{s(f)}}{d_{s(f)}}/{D_{s(f)}}}}{{{\sigma _s}{d_s}/{D_s} + {\sigma _f}{d_f}/{D_f}}}, {D_e} = {D_s}{\eta _s} + {D_f}{\eta _f}, \\
{I_e} = I{\eta _f},{\Delta _e} = \Delta_{s} {\eta _s}, \\ 
\end{array}
\end{equation}
  where $d_{s(f)}$ is the thickness of the superconducting (ferromagnetic) layer and $\sigma _{s(f)}$ is the transparency parameter on the side of the superconducting (ferromagnetic) layer. Comparing the phase diagrams of a uniform magnetic superconductor with the effective parameters\,(\ref{effective}) and the phase diagram of the superconductor–ferromagnet system, we conclude that approximation (\ref{effective}) describes satisfactorily the dependence of the critical temperature on the thickness of a ferromagnet in the thickness range $d_{f}<d_{f0}$, where
$d_{f0}$ is the thickness of the ferromagnet corresponding to the first minimum of the dependence $T_c(d_f)$. In the case of a superconductor contacting a uniform iron layer, the thickness of 8\,\AA\ corresponds to a thickness of approximately $d_{f0}$. Thus, this approximation is suitable for estimating the effect of small-scale magnetic inhomogeneities in a given system. To calculate the critical temperature of the S/F system with a planar domain structure in approximation (\ref{effective}), we can use the self-consistency equation (\ref{UE})) for the Usadel matrix function $\hat F$ 
\begin{equation}
\label{UE}
\begin{array}{l} 
{\Delta _e}\ln \left( {\frac{{{T_c}}}{{{T_{cs}}}}} \right) = \pi {T_c}\sum\limits_{\omega  >  0}^{{\omega _D}} {\left( {\Sp{\hat F}\left( {x,\omega } \right) - \frac{{{\Delta _e}}}{\omega }} \right),}  \\ 
\end{array}
\end{equation}
 where $\omega$ is the Matsubara frequency.
 
Using the algorithm for solving the boundary value problem for the Usadel matrix function \cite{bib:Tumanov16}, we found that the effect of the ferromagnetic exchange interaction on the critical temperature is almost completely suppressed for a characteristic size of the domain structure (1\thinspace-\thinspace2)$\xi_s$. In this case, it is possible to compare the given characteristic size of the magnetic inhomogeneities with the effective exchange field $I_e$ (acting in a homogeneous magnetic superconductor), which leads to the same decrease in $T_c$. We note that this approximation is suitable for estimating the critical temperature, but is not suitable for studying the transport properties of the system. 

We calculated the critical temperature of the superconducting transition for the (Fe/Cr/Fe)/V/Fe systems in the single-mode approximation, taking into
account various realized variants of magnetic ordering in the Fe/Cr/Fe structure (see Fig.\,\ref{Tc}). The dependence of the magnetization on the thickness of the chromium layer was estimated within the phenomenological model of chromium deposition on iron (see Fig.\,\ref{GR}), taking into account the varying degree of filling of the upper iron layer. In addition, we used a phenomenological estimate of the long-range RKKY-type contribution to the interaction of magnetizations in iron layers separated by chromium, which was obtained by separating changes in the mutual orientation of the magnetization that are incommensurate with the lattice constant of chromium. Such estimates are in good agreement with the theoretical calculations of the magnetic interaction of iron layers separated by a chromium layer in similar systems \cite{bib:Wang90}. To determine the parameters of the phenomenological model, we compared our experimental data and the data of \cite{bib:Unguris_PRL}, where the mutual orientation of the magnetization of the iron layers was measured directly.

 \begin{figure}
	\centering
	\includegraphics[width=0.97 \linewidth]{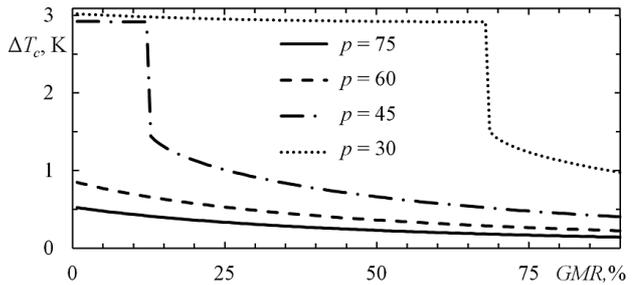}
	\caption{\textbf{Fig.\,4.} Difference in the critical temperatures of the
		Fe(4\AA)/Cr/Fe(4\AA)/V(125\AA) system with antiparallel and
		parallel oriented magnetizations of the iron layers separated
		by a chromium layer.}
	\label{SW}    
\end{figure}

 A set of parameters obtained by comparing the theory with our experiment can be used to estimate the critical temperature of such a system with other layer thicknesses. Using the resulting parameters, we calculated the critical temperature of the Fe(4\,\AA)/Cr/Fe(4\,\AA)/V(125\,\AA) system with parallel $T_{c(p)}$ and antiparallel $T_{c(ap)}$ magnetizations of iron layers separated by the chromium layer. The dependence of the difference of these critical temperatures $\Delta T_c = T_{c(ap)} - T_{c(p)}$ on the magnitude of the giant magnetoresistance effect in the Fe/Cr/Fe system is presented in Fig.\,\ref{SW}. The parameter $p$ characterizing the transparency of the Fe/Cr interfaces significantly affects the magnitude of the effect. When comparing our theory with experiment, the best agreement can be achieved in the range $p = 30 \mathrm{-} 60$ depending on the thickness of the chromium layer. We note that the magnitude of the spin valve effect in this geometry is very sensitive to the transparency of the F/S interface and the thickness of the superconducting layer.
 
  To summarize, the magnetic and superconducting properties of thin-layer Fe/V/Fe and (Fe/Cr/Fe)/V/Fe systems have been experimentally investigated. The magnetic properties of the (Fe/Cr/Fe)/V/Fe system have been studied by FMR. Qualitative changes in the shape of the FMR line after the introduction of a chromium layer have been observed, beginning with an effective thickness of less than one monolayer. The superconducting transition temperature as a function of the thickness of the chromium layer in the (Fe/Cr/Fe)/V/Fe system has been determined from the jump in the magnetic susceptibility in wedge-shaped samples. Within the theory of the proximity effect in the dirty limit, we have calculated the critical temperature as a function of the thickness of the chromium layer. Our theoretical estimates are in semiquantitative agreement with the experimental data on the critical temperature of the system as a function of the thickness of chromium. According to our estimates, sections of samples with a critical temperature
of 3.8--4.2\,K correspond to situations where the iron layer between chromium and vanadium is
split into domains whose characteristic size is about the superconducting coherence length. Observation of the critical temperature below 2\,K indicates a relatively low transparency of the interface between iron and chromium. Using the resulting parameters, we have estimated the effect of the superconducting spin valve in the Fe(4\,\AA)/Cr/Fe(4\,\AA)/V(125\,\AA) system. The difference
in critical temperatures for antiparallel and
parallel orientations of the magnetization of iron layers
can reach 1--2\,K. The results obtained can be useful
in the design of a superconducting spin valve \cite{bib:Kushnir16}.

This work was supported in part by the Ministry of
Education and Science of the Russian Federation
(subsidy allocated to the Kazan Federal University for
project no. 3.2166.2017 within the state research
assignment). The work of V.A.T. was also supported by
the Russian Foundation for Basic Research (project
no. 16-02-01016).

\end{document}